\documentclass[pra,twocolumn,showpacs,preprintnumbers,
amsmath,amssymb,floatfix]{revtex4}
\usepackage{hyperref}
\usepackage[dvips]{epsfig}
\newcommand{\beq}{\begin{equation}}
\newcommand{\eeq}{\end{equation}} 
\newcommand{\beqa}{\begin{eqnarray}}
\newcommand{\eeqa}{\end{eqnarray}}
\newcommand{\ba}{\begin{array}}
\newcommand{\ea}{\end{array}}

\begin{document}

\title{Shock waves in strongly interacting Fermi gas \\
from time-dependent density functional calculations}
\author{F. Ancilotto$^{1,2}$, L. Salasnich$^{1}$, and F. Toigo$^{1,2}$} 
\affiliation{$^1$Dipartimento di Fisica e Astronomia 
"Galileo Galilei" and CNISM, Universit\`a di Padova, 
Via Marzolo 8, 35122 Padova, Italy \\
$^2$CNR-IOM Democritos, via Bonomea, 265 - 34136 Trieste, Italy} 

\begin{abstract} 
Motivated by a recent experiment [Phys. Rev. Lett. {\bf 106}, 150401 (2011)]
we simulate the collision between two clouds of cold Fermi gas at
unitarity conditions by using an extended Thomas-Fermi 
density functional. At variance with the current interpretation of
the experiments, where the role of viscosity is
emphasized, we find that a quantitative agreement
with the experimental observation of the dynamics of the 
cloud collisions is obtained within our superfluid 
effective hydrodynamics approach, where density 
variations during the collision are controlled by a purely dispersive 
quantum gradient term. We also suggest different initial conditions 
where dispersive density ripples can be detected with the available 
experimental spatial resolution. 
\end{abstract} 


\maketitle

A Fermi gas of atoms at unitary conditions, 
i.e. when the s-wave scattering length diverges \cite{bertsch}, 
is predicted to obey universal hydrodynamics, where the shear viscosity and
other transport coefficients are universal functions of 
density and temperature.
Experiments on the expansion of rotating, strongly interacting
Fermi gas in the normal fluid regime 
reveal extremely low viscosity hydrodynamics \cite{cao},
thus making a strongly interacting Fermi gas in the normal fluid
regime a nearly perfect fluid with almost frictionless mass flow.
Remarkably, the superfluid behaves in rotating clouds experiments almost
identically to the normal fluid one. Consistently with such observations, 
theory predicts \cite{son1} that the unitary Fermi gas
has zero bulk viscosity in the normal phase, whereas
in the superfluid phase two of the three bulk viscosities 
vanish \cite{nota,sommer}.

Nonlinear evolution of trapped cold gases is often 
characterized by the appearence of wave-like localized
distorsion of the trapped gas cloud.
In addition to the well-known 
sound waves, shock waves often appear, which are characterized by 
an abrupt change in the density of the medium \cite{landau,whitham}.
Shock waves are ubiquitous and 
have been studied in many different physical systems \cite{landau,whitham}. 
Very recently the observation of nonlinear hydrodynamic waves 
has been reported in the collision between two strongly 
interacting Fermi gas clouds of $^6$Li atoms 
at unitarity \cite{thomas}. 
During the collision, which has been experimentally 
imaged in real time \cite{thomas}, 
when regions of high density move with 
faster local velocity than region of low 
density, large density gradients develop 
when the two clouds merge with each other.
Unlike the case of collision between 
two initially separated BEC in a 
cigar shaped trap, where the shock waves were 
interpreted as purely dispersive \cite{chang},
the authors of Ref. \cite{thomas} invoke  dissipative forces to avoid 
the "gradient catastrophe"  in the unitary Fermi system.
To describe their experimental data, they use  
an effective one-dimensional (1D) model 
based on
time-dependent non-linear hydrodynamical equations, with
a phenomenological kinematic viscosity term added to describe
dissipative effects, whose strength 
is used as a fitting parameter. 
A quite high value of the viscosity coefficient
is necessary in the description of the 
collision in order to reproduce the experimental data.
The origin of such viscosity, however, 
is not clarified in Ref. \cite{thomas}.
In fact the role of dissipation itself in cold Fermi gas at low
temperature is questionable, since the ultracold
unitary Fermi gas is known as an example 
of an almost perfect fluid, as discussed above. 
In this paper we offer an alternative explanation, where 
the collision process is dominated by dispersion effects \cite{disp}.
Evidence supporting the notion that 
shock waves in ultracold Fermi gas at unitarity are 
dispersive rather than dissipative can be found in 
the simulations reported in Ref.\cite{new-bulgac},
based on the self-consistent solution of coupled 
Bogoliubov-de Gennes equations 
derived from the the zero-temperature time-dependent 
superfluid local density approximation \cite{bulgac}.
Our calculations are based on a single-orbital density functional 
(DF) approach to the properties of unitary Fermi gas at low temperatures, 
which has been used recently to successfully 
address a number of properties of such system 
\cite{salasnich,best,anci1,anci2,salasnich-ratio,salasnich-shock}. 
The advantage in using a single-orbital DF approach is that 
systems with a very large number of particles 
can be treated using only a single function of the coordinate, 
i.e. the particle density. Thus, it might represent a viable
alternative to the much more computationally 
expensive approaches (like, e.g., the Bogoliubov-de Gennes method) 
often used to describe superfluid Fermions. 
This is especially true in the 
case of 3D geometries, like the ones investigated here,
where the BdG method would be prohibitively costly. 

In our extended Thomas-Fermi (ETF) density functional 
approach \cite{salasnich,best} the total energy of the unitary Fermi gas 
is given by 
\beq 
E[n] = \int d^3{\bf r} \ {\cal E}(n,{\boldsymbol \nabla} n) \; 
\label{e-dft}
\eeq
where 
\beq 
{\cal E}(n,{\boldsymbol \nabla} n) = 
\lambda {\hbar^2 \over 8 m} {(\nabla n)^2\over n} + 
{ \xi} {3\over 5} {\hbar^2 \over 2m} (3\pi^2)^{5/3} n^{5/3}
+ U({\bf r}) \, n   \; .  
\label{ee-dft}
\eeq 
Here $n({\bf r})$ is the fermions number 
density and $U({\bf r})$ is the confining external potential.
The total energy functional ${\cal E}$ 
contains a term proportional to the kinetic
energy of a uniform non interacting fermions gas, plus 
a gradient correction of the von-Weizsacker form \cite{von}.  
In recent papers \cite{salasnich,best,anci1,salasnich-ratio} 
we have determined the parameters $\xi$ and $\lambda$
by fitting Monte Carlo results 
\cite{blume,thermo-bulgac} for the energy of fermions confined in 
a spherical harmonic trap close to unitary conditions.
The main conclusion of that work is that the values 
\beq 
\xi = 0.40 \quad\quad \mbox{and} \quad\quad \lambda = 1/4 \;    
\eeq
fit quite well Monte Carlo data of the unitary Fermi gas.  
In particular the chosen value for 
${ \xi}$ almost coincides with the experimental 
determination of Ref. \cite{turlapov}.

The density functional (\ref{e-dft}) describes 
various static and dynamical properties of the unitary Fermi 
gas trapped by an external potential. 
The gradient term in the previous equation is found to be 
crucial to describe accurately the surface effects of 
the system, in particular in systems with a small number of atoms, 
where the Thomas-Fermi (local density) approximation 
fails \cite{salasnich,anci1}. 
As a key result of the present paper, we show  
that when fast dynamical processes occur and/or when 
shock waves come into play such term is necessary 
also in the large N limit.
 
The extended non-viscous and irrotational 
hydrodynamics equations deriving from the 
functional (\ref{e-dft}) are given by 
\beqa
{\partial n \over \partial t} + {\boldsymbol 
\nabla} \cdot (n {\bf v}) = 0 \; , 
\label{hy1}
\\
m{\partial {\bf v} \over \partial t} + {\boldsymbol 
\nabla} [ {m\over 2} v^2 + 
{\partial {\cal E}\over \partial n}- {\boldsymbol \nabla}\cdot 
{\partial {\cal E}\over \partial ({\boldsymbol \nabla} n)}] = {\bf 0} \; ,  
\label{hy2}
\eeqa
where $n({\bf r},t)$ is the time-dependent scalar density field 
and ${\bf v}({\bf r},t)$ the time-dependent vector velocity field.  
If $\lambda =0$, then Eqs. (\ref{hy1}) and (\ref{hy2}) reproduce  the 
equations of superfluid hydrodynamics \cite{stringa-fermi} by construction.

Notice that equations (\ref{hy1}) and (\ref{hy2}) 
can equivalently be written in terms of a superfluid time-dependent 
nonlinear Schr\"odinger equation (NLSE) 
involving a complex order parameter \cite{salasnich}. 
We will numerically solve this NLSE equation to obtain the long-time 
dynamics of the collision between two 
initially separated Fermi clouds, as described in
the following. Our goal is to simulate 
the experiments of Ref. \cite{thomas}. We have used 
the Runge-Kutta-Gill fourth-order method \cite{rkg} 
to propagate in time the solutions of the NLSE. 
To accurately compute the spatial derivatives 
appearing in the NLSE, we used a 13-point finite-difference formula 
\cite{pi}. 

Since the confining potential used in the experiments
is cigar-shaped, we have exploited the resulting cylindrical symmetry 
of the system by representing the solution of our NLSE on a 2-dimensional
$(r,z)$ grid (of 500$\times $2500 uniformly spaced points). 
Of course, this choice, which greatly reduces the computational 
cost of the simulations, would not be able to describe 
possible, azimuthal dependent, transverse instabilities and 
vortex formation (like those observed in 
the collision between BEC clouds \cite{chang}). 
Although such features are apparently 
not observed in the collision experiments
of Ref.\cite{thomas},
we cannot rule out the possibility of  
transverse instabilities 
and vortex formation in Fermi gas clouds for different 
initial conditions than those investigated here. 
To describe the details of such structures,  
a full 3-D simulation is needed.  

In our simulations we tried to reproduce as closely as 
possible the experimental conditions of Ref. \cite{thomas},
which we summarize briefly in the following. 
A 50:50 mixture of the two lowest hyperfine states of $^6$Li
is confined by an axially symmetric cigar-shaped laser trap,
elongated along the z-axis. The resulting trapping potential is  
$U(r,z)=0.5m[\omega _r^2r^2+ \omega _z^2z^2]$,
with $\omega _r=2\pi \times 437$ Hz and $\omega _z=2\pi \times 27.7$ Hz.
The trapped Fermi cloud is initially 
bisected by a blue-detuned beam which provides a
repulsive knife-shaped potential.
This potential is then suddenly removed, allowing for the
two separated part of the cloud to collide with each other. The system is
let to evolve for a given hold time
$t$, then the trap is removed in the radial direction,
and the system is allowed to evolve for 
another $1.5$ ms during which the gas expands in the r-direction
(during this extra expansion time, the confining trap frequency along
the z-axis is changed to $\omega _z=2\pi \times 20.4$ Hz),
and finally a (destructive) image of the cloud
is taken. The process is repeated from the beginning for 
another different value for the hold time $t$. The experimental results are
eventually plotted as 1D integrated density profiles
for the different hold times investigated: in Fig. 2 of Ref. \cite{thomas},  
the experimental 1D density profiles at different 
times $t$ are shown along the long trap axis. 

We simulated the whole procedure within the framework discussed
above. As in  Ref. \cite{thomas}, we choose
the initial density profile 
in the form of a static solution of  hydrodynamic equations:
\beq
n (r,z,t=0) = \tilde{n}(1-{r^2 \over R^2}-{z^2\over R_z^2}-
{V_{rep}(z) \over \mu _G})^{3/2}
\label{dens}
\eeq
where $\tilde{n}=[(2m\mu _G/\hbar ^2)/\xi ]^{3/2}/(3\pi ^2)$.
$V_{rep}$ represents an optically generated 
knife-shaped repulsive potential
used to initially split the Fermi cloud into two
spatially separated components that are led to 
collide with each other upon removal of such potential.
$V_{rep}=V_0exp(-(z-z_0)^2/\sigma _z^2)$,
where $V_0=12.7\,\mu K$, 
$\sigma _z=21.1\,\mu m$ and $z_0=5\,\mu m$. 
In Eq.\ref{dens} we use the same values as in Ref.\cite{thomas},
which provide a fit to the observed initial experimental cloud 
density profile immediately after the
removal of the knife potential.
Here $R_z=220\, \mu m$ and $R=14 \,\mu m$.
In particular, the chosen value for the chemical potential
$\mu _G=0.53\,\mu K$
corresponds to a total of $N=2\times 10^5$ $^6$Li atoms.

During the time evolution of our system, 
when the two clouds start to overlap, many ripples 
whose wavelength is comparable to the interparticle distance
are produced in the region of overlapping densities. These ripples 
also propagate backwards towards the trap boundaries,
affecting larger and larger portion of the simulated cloud.
We stress that these effects are quite similar to those 
found in a recent experiment \cite{chang} by merging and 
splitting Bose-Einstein condensates. 

In order to properly compare our results with the 
experimental data of resonant fermions \cite{thomas}, 
which are characterized by a finite
spatial resolution, we smooth the calculated profiles at 
each time $t$ by local averaging the density within a space window 
of $\pm 5$ $\mu m$ centered around the calculated point.
This procedure will give smoothed density profiles 
with a spatial resolution close to the one 
characterizing the experimental setup of Ref. \cite{thomas}.

\begin{figure}
\centerline{\psfig{file=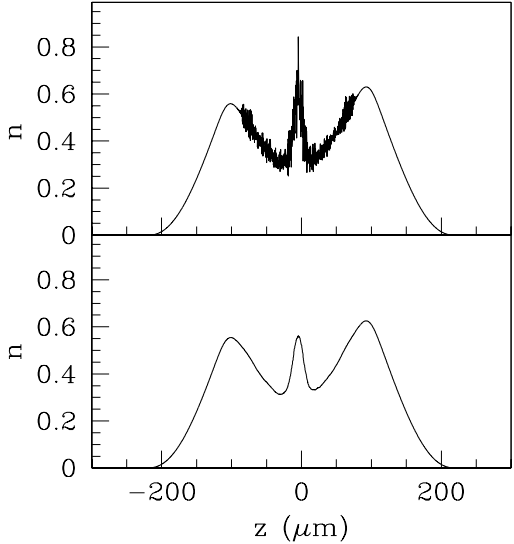,height=4.in,clip=}}
\caption{ Density profile after t=$1.5\, ms$. The two curves
show the smoothed and non-smoothed calculated profiles,
respectively.
The normalized density is in units of $10^{-2}/\mu m$ per particle.}
\label{fig1}
\end{figure}

The effect of smoothing is 
shown in Fig.\ref{fig1}, where the simulated 
density profile 
during the time evolution of the colliding clouds
is shown before and
after the local averaging procedure is applied.
We wish to stress that this smoothing procedure
is just a post-processing of the data obtained by the 
time evolution of the  NLSE corresponding to Eqs.  (\ref{hy1}) and (\ref{hy2})
and therefore it does not affect the time evolution itself.

\begin{figure}
\epsfig{file=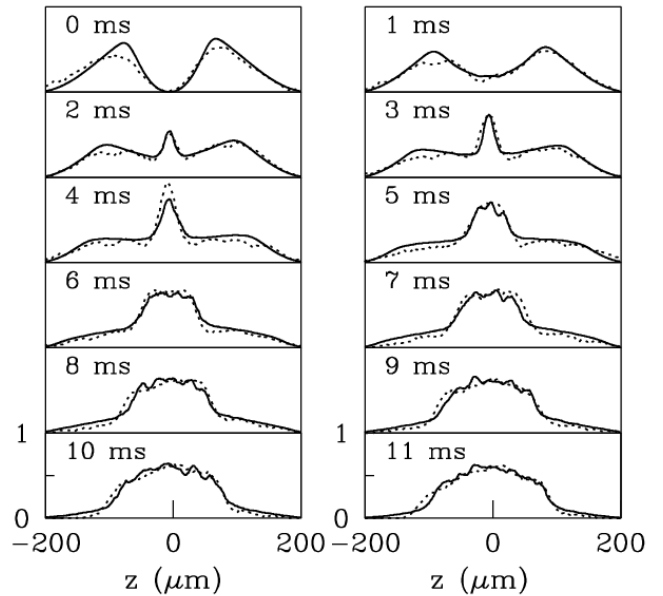,height=3.3 in,clip=}  
\caption{1D density profiles at different times $t$ 
showing the collision of two strongly interacting
Fermi clouds. Solid lines: our calculations 
with no adjustable parameter. 
Dotted lines: experimental data from Ref. \cite{thomas}. 
The normalized density is in units of $10^{-2}/\mu m$ per particle.}
\label{fig2}
\end{figure}

The results of our simulations (solid lines), for the whole
time duration of the experiments, and after 
the smoothing procedure is applied to the (y-averaged) density 
profile at each time, are shown in Fig. \ref{fig2} 
plotted along the long trap axis, for the same
time frames as in the experiment. 
The experimental results (dotted lines) are also shown for comparison.

The time value shown in each frame corresponds to the time 
evolution of the initial profile, Eq. (\ref{dens}), before 
the trap in the radial direction is removed to let 
the system evolve for another $1.5$ $ms$ in the axial trap only, 
as done in the experiment.

Note the striking correspondence between the experimental data 
and our simulation. 
We emphasize the fact that our simulations do not have adjustable
parameter to be used to fit the experimental 
data, at variance with the model calculations
presented in Ref. \cite{thomas}. 
Even after the smoothing procedures is applied,
our simulated density profiles exibit 
short-scale structures superimposed to the body of the 
cloud profiles. Such ripples are indeed present 
also in the experimental data, whereas in the
model profiles used in Ref. \cite{thomas} to fit the observed images 
such oscillations are completely absent, due to the presence of 
a strong viscosity term in their model.

The numerical results shown in Fig. \ref{fig2} have been obtained 
by using $\lambda =1/4$, as previously discussed. 
In order to check the robustness of our results upon
a different choice for $\lambda $, we have performed
various time-dependent calculations with the same initial
conditions but using different values for $\lambda$.
It turned out that changing $\lambda $ from the
optimal value $\lambda =1/4$ has profound consequencies on the
long-time evolution of the colliding clouds, 
providing density profiles which are completely 
different from the experimental ones.

\begin{figure}
\epsfig{file=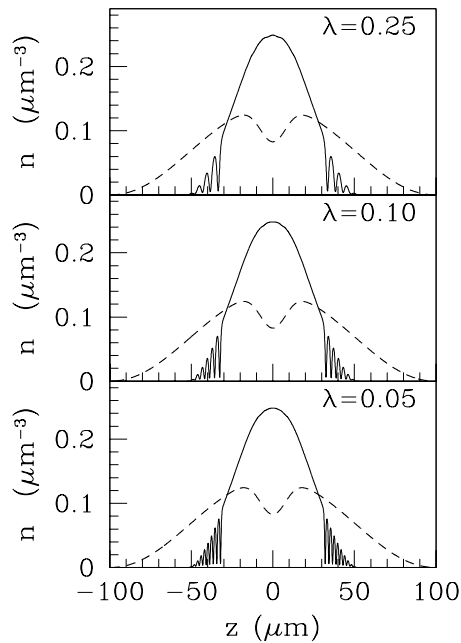,height=4. in,clip=}  
\caption{"Soft" collision. The initial density profile is
shown with dashed lines.} 
\label{fig3}
\end{figure}
This finding is not simply a numerical result, but has an 
important physical bearing.
In fact, a strong dependence on $\lambda$ of the time evolution of 
a Fermi cloud made of a large number of atoms
is at first sight surprising, given the fact that
the gradient term should become less and less
important with increasing  $N$.
We believe that such dependence is due to the 
presence of shock waves (i.e. regions characterized by
large density gradients) in the colliding clouds.
To check this hypothesis one should perform experiments 
with different initial conditions, which do not evolve 
into shock waves.
As an example, we have simulated  a "soft" collision, 
where shock waves are not expected,
by considering a system 
with a smaller density of fermions ($N=4000$ in the same trap 
used in the simulations of Fig. \ref{fig2}) and where
the two initial clouds are largely overlapping at the
beginning of the simulation (this could be obtained experimentally
by reducing the heigth/width of the "knife potential"),
thus reducing the velocity of the impinging
clouds. It turns out that the long-time behavior of
such system is indeed {\it independent} on the chosen value 
of $\lambda$. However, even for such a system, 
shock waves will eventually occur.  As expected, such waves 
break into ripples whose wavelength 
depends on $\lambda$. 
This is illustrated in Fig. \ref{fig3}, where the initial overlapping 
clouds are shown, together with the profile after $t=10\, ms$.
For $t < t_s \simeq 9\, ms$ the shape of the time-evolved 
cloud is exactly the same for all values of $\lambda$. 
After the shock time $t_s$, however, the density profile develops a steep 
density gradient at the cloud boundaries, 
visible in the figure, which eventually 
breaks into a train of ripples. The three images shown in Fig.\ref{fig3}
are taken just after the ripples have been produced. 

\begin{figure}
\epsfig{file=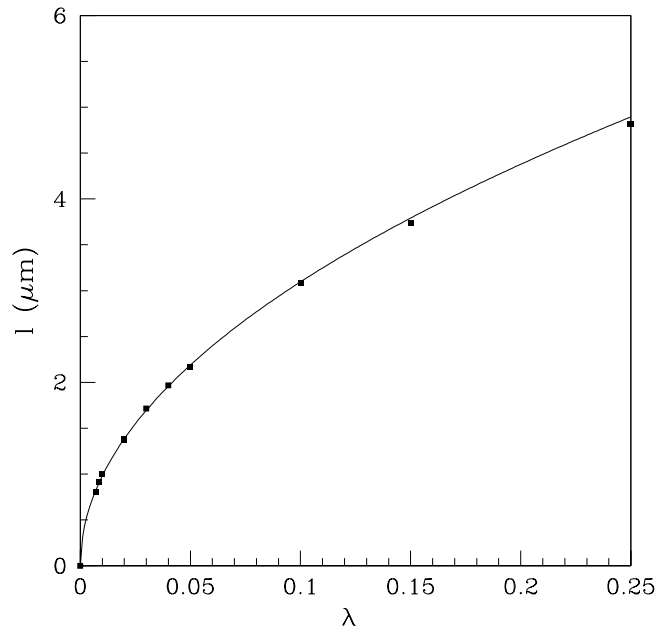,height=3.5in,clip=}  
\caption{ Ripples wavelength $l$ as a function of gradient 
coefficient $\lambda$ of the density functional. 
The solid line shows the $\sqrt{\lambda}$.  } 
\label{fig4}
\end{figure}

These ripples are  characterized by 
a well-defined wavelength $l$, which we estimated from the
computed profiles. 
We plot in Fig. \ref{fig4} the calculated ripples wavelength $l$
as a function of $\lambda$. We note that the calculated values for $l$ 
follows very closely a $\sqrt {\lambda }$ law, which is expected 
on the basis of dimensional analysis by balancing the energy 
associated to dispersion to that of the nonlinearity.

In conclusion, we have numerically studied the
long-time dynamics of shock waves in the ultracold unitary Fermi gas.
We have described the system by using an extended density 
functional approach, which has been used recently to
successfully describe a number of static and dynamical 
properties of cold Fermi gases. 
Two main results emerge from our calculations:
a) at zero temperature the simplest regularization 
process of the shock is purely dispersive, mediated by the 
quantum gradient term, 
which is one of the ingredient in our DF approach;
b) the quantum gradient term plays an important role 
not only in determining the static density profile of 
small systems, where surface effects are important, 
but also in the fast dynamics of large systems, 
where large density gradients may arise.

Finally, we stress that 
dispersive shock waves with a characteristic wavelength 
should be observable, according 
to our simulations, by using a soft-collision setup. 

We thank James Joseph and John E. Thomas for useful 
comments and suggestions, and for having sent us 
their experimental data.


\begin{thebibliography}{99}
\bibitem{nota} Whereas mass flow in unitary Fermi gas 
is essentially frictionless, 
the opposite is true for spin flow, where maximum 
resistence to flow is observed \cite{sommer}: as
a consequence 
two clouds of ultracold fermions prepared in different spin states
bounce off each other, in spite of the fact that the underlying 
atomic interaction is attractive and the equilibrium 
ground state is known to be a paired superfluid.
\end{thebibliography}
\end{document}